\documentstyle[prb,twocolumn,epsf,aps]{revtex}
\begin{document}
\draft \preprint{} \twocolumn[\hsize\textwidth\columnwidth\hsize\csname
@twocolumnfalse\endcsname
\title{Injection statistics simulator for dynamic analysis of noise in
mesoscopic devices}
\author{T. Gonz\'alez,
J. Mateos, and D. Pardo}
\address{
Departamento de F\'{\i}sica Aplicada, Universidad de Salamanca, Plaza de
la Merced s/n, E-37008 Salamanca, Spain}
\author{L. Varani}
\address{
Centre d'Electronique et de Micro-opto\'electronique de Montpellier (CNRS
UMR 5507), Universit\'e Montpellier II \\ 34095 Montpellier Cedex 5,
France}
\author{L. Reggiani}
\address{
Istituto Nazionale di Fisica della Materia, Dipartimento di Scienza dei
Materiali, Universit\`a di Lecce \\ Via Arnesano, 73100 Lecce, Italy}
\date{\today}
\maketitle
\begin{abstract}
{We present a model for electron injection from thermal reservoirs which
is applied to particle simulations of one-dimensional mesoscopic
conductors. The statistics of injected carriers is correctly described
from nondegenerate to completely degenerate conditions. The model is
validated by comparing Monte Carlo simulations with existing analytical
results for the case of ballistic conductors. An excellent agreement is
found for average and noise characteristics, in particular, the
fundamental unities of electrical and thermal conductances are exactly
reproduced.}
\end{abstract}
 \vskip1pc]
\narrowtext
The systematic trend of reduction in the size of electronic devices has
led to the appearance of new phenomena that require special attention to
be properly investigated. In particular, mesoscopic conductors are
attracting increasing interest in recent years.\cite{kogan96,dejong97}
Here, the microscopic interpretation of carrier transport and fluctuations
demands for approaches which differ from those typically used in
macroscopic devices. Several techniques have been used to this end.
Accordingly, to account for phase coherence the scattering matrix theory
originally proposed by Landauer \cite{landauer57} has been further
elaborated.\cite{buttiker90,martin92,buttiker92} A Wigner function
formalism has also been used for the analysis of ballistic and diffusive
conductors.\cite{kuhn92a,kuhn92b,reggiani95,greiner97} When phase
coherence does not play an essential role, semiclassical methods based on
the Boltzmann-Langevin equation have shown to provide viable
solutions.\cite{nagaev92,naveh97,suk98} In all these theoretical
approaches, the modeling of the contacts is proven to be crucial. The
active region of the devices is considered to be surrounded by leads which
are usually treated as ideal thermal reservoirs. In other words, contacts
are assumed to be always at thermal equilibrium; absorbed carriers are
thermalized immediately, thus any correlation is destroyed, and emitted
carriers obey a Fermi-Dirac distribution.
\par
Recently, particle simulations, mainly based on the Monte Carlo method,
have been used to study fluctuations in mesoscopic
structures.\cite{liu97,shot1,shot3,shot4,shot5} This technique, which is
widespread for the analysis of macroscopic electronic
devices,\cite{jacoboni89,varani94} has the advantage of being applicable
under physical conditions which can be very far from thermal equilibrium,
and often can not be studied analytically. Moreover, it can provide
detailed microscopic information about the physical processes and the time
scales associated with transport and fluctuations in electronic devices.
This last feature makes particle simulations quite attractive for the
study of mesoscopic conductors. In the case of macroscopic devices, the
presence of energy dissipation and diffusive regions inside the structures
washes out the influence of the contact injecting statistics on the output
currents and voltages, and the simulation of contacts does not require
very detailed models.\cite{woolard94,gonzalez96} On the contrary, when
dealing with mesoscopic structures, the modeling of carrier injection from
thermal reservoirs is a delicate problem. In particular, the statistics of
electron injection associated with a Fermi-Dirac distribution at the
electrodes is essential for the correct analysis of fluctuations and
effects related to Pauli exclusion principle. For a classical injector,
the statistics of transmitted charge is Poissonian,\cite{kogan96} and can
be easily accounted for.\cite{shot1} On the contrary, under degenerate
conditions one should use a binomial distribution,\cite{levitov93} and to
our knowledge this issue has not been addressed so far.
\par
The aim of this paper is to present a model for particle injection from
ideal thermal reservoirs into one-dimensional mesoscopic conductors which
takes into account the fluctuating occupancy of the incoming electron
states associated with a Fermi-Dirac distribution. The model can be
continuously applied from nondegenerate to degenerate statistics at the
contacts. The results obtained with a Monte Carlo simulation of a
one-dimensional two-terminal ballistic conductor implementing the present
contact model are compared with existing analytical results to validate
the injection scheme.
\par
Let us consider a one-dimensional conductor connected to leads which act
as perfect thermal reservoirs. The density (in $k$-space) of incoming
electron states with wave vector $k$ impinging per unit time upon the
boundary between the leads and the conductor, $\zeta_k$, is given by the
product of the density of states $n_k$ and the velocity $v_k$ normal to
the boundary, $\zeta_k=n_kv_k={1 \over \pi}{\hbar k \over m}$, where we
have taken a parabolic isotropic $\varepsilon-k$ relation. These $n_k$
states obey Fermi-Dirac statistics, thus only a fraction
$f(\varepsilon_k)=\{1+\exp[(\varepsilon_k-\varepsilon_F)/k_BT]\}^{-1}$ of
them will be occupied and eventually will inject a carrier into the
conductor, with $\varepsilon_F$ the Fermi level. Therefore, the injection
rate density of carriers with momentum $k$, $\Gamma_k$, is given by
$\Gamma_k=\zeta_kf(\varepsilon_k)$. While $\zeta_k$ does not depend on
time, the instantaneous occupancy of an incoming $k$-state
$\widetilde{f}(\varepsilon_k,t)$, of which $f(\varepsilon_k)$ is the
average, fluctuates in time obeying a binomial
distribution\cite{levitov93} with a probability of success
$f(\varepsilon_k)$. The injecting statistics imposed by this binomial
distribution is determined by the Fermi-Dirac statistics electrons obey,
i.e., ultimately, by Pauli principle. When $\varepsilon_k-\varepsilon_F
\ll -k_BT$, $f(\varepsilon_k)\cong 1$ and the injecting statistics of the
corresponding $k$-state becomes uniform in time. On the contrary, when
$\varepsilon_k-\varepsilon_F \gg k_BT$, $f(\varepsilon_k)\ll 1$ and the
injecting statistics of the corresponding $k$-state becomes Poissonian in
time. In a completely degenerate (quantum) reservoir, the former
condition is fulfilled for any incoming $k$-state and the injection is
uniform in time for all the $k$ values up to the Fermi wave vector $k_F$.
On the contrary, in a nondegenerate (classical) reservoir, the latter
condition applies for all $k$ values.
\par
To reproduce the injecting statistics imposed by the Pauli principle in a
particle simulation, it is necessary to discretize momentum space into a
certain number of meshes of width $\Delta k$ around discrete values of
$k$, $k_i$. For each of these meshes, the number of incoming electron
states per unit time with wave vector $k_i$ is given by $\zeta_{k_i,\Delta
k}=\zeta_{k_i} \Delta k$, with a probability of occupancy given by
$f(\varepsilon_{k_i})$. In the simulation, at each time interval of
duration $1 / \zeta_{k_i,\Delta k}$ an attempt to introduce an incoming
electron with wave vector $k_i$ takes place. At this point a random number
$r$ uniformly distributed between 0 and 1 is generated, and the attempt is
considered successful only if $r<f(\varepsilon_{k_i})$ . This
rejection-technique scheme properly accounts for the injection statistics
at each mesh in $k$-space.
\par
For a completely degenerate reservoir, in every mesh up to $k_F$ an
electron is injected every time interval $1 / \zeta_{k_i,\Delta
k}$,\cite{remark1} and there is no need of the rejection technique. This
is the case of the simple contact modeling used in Ref.
\onlinecite{liu97}. For a nondegenerate reservoir, since
$f(\varepsilon_{k_i})\ll 1$ for all $k_i$-states, it is possible to use a
global Poissonian statistics characterized by an injection rate
$\Gamma_{clas}=\int_0^{+\infty}\Gamma_kdk$. Accordingly, the time between
two consecutive electron injections is generated with a probability per
unit time $P(t)=\Gamma_{clas} e^{-\Gamma_{clas} t}$. Then, the electron
wave-vector is randomly picked from a Maxwell-Boltzmann distribution, and
there is no need of using a mesh in $k$-space. This is the scheme used in
Refs. \onlinecite{shot1,shot3,shot4,shot5}. For any intermediate level of
degeneracy, to account for the proper statistics at each value of $k_i$,
it is necessary to use the scheme explained above, which of course is also
valid in the classical and degenerate limits, but less efficient from the
point of view of computation time. The accuracy of the proposed scheme
depends on the number of meshes in $k$-space used to inject carriers. In
any case, it is not necessary to use a very large number of meshes.
Indeed, it is well known that the noise of an electrical system depends
only on the kinetics of electron states in a small energy range around the
Fermi level. We have checked that very satisfactory noise results can be
obtained by simulating just an energy range of $3k_BT$ above and below
$\varepsilon_F$, and dividing it into $50$ meshes. Below and above this
range all $k$-states can be respectively considered to be completely
occupied and empty. Therefore, they do not contribute to current
fluctuations and can be ignored.
\par
In the following we will report the results of a Monte Carlo simulation of
a one-dimensional two-terminal ballistic conductor of length $L$ connected
with two thermal reservoirs modeled according to the above scheme. The
temperature is taken to be 300 K and the effective mass $m=0.25 \ m_0$,
$m_0$ being the free electron mass. Carriers are considered to move
ballistically into the conductor following the classical equations of
motion, and when a voltage $U$ is applied to the leads, electrons are
accelerated by an electric field $E=U/L$. When a carrier inside the
conductor reaches a contact, it is considered to be immediately
thermalized and it is cancelled from the simulation. For simplicity, the
cross-sectional area of the conductor is assumed to be sufficiently small
so that only the lowest sub-band is occupied. We remark that Coulomb
interactions are ignored and Pauli exclusion principle is taken into
account only at the contact injection. Under these conditions, the
literature provides several analytical results which will be used to
validate the model.
\par
Figure~\ref{sieq} shows the low-frequency value of the current spectral
density $S_I(0)$ at equilibrium normalized to $2qI_S$ as a function of the
degeneracy factor $\varepsilon_F/k_BT$, with $\varepsilon_F$ measured with
respect to the bottom of the conduction band.
$I_S=q\int_0^{\infty}v_kn_kf(\varepsilon_k)dk$ is the saturation current,
i.e., the maximum current a contact can provide. In the classical limit,
corresponding to large negative values of the degeneracy factor,
$S_I(0)=4qI_S$. Here all carriers contribute to the current noise and
$S_I(0)$ is just the sum of the full shot noise related to the two
opposing currents $I_S$ injected by the
contacts.\cite{kuhn92a,reggiani95,shot1,shot5} Under degenerate
conditions, corresponding to positive values of the degeneracy factor,
$S_I(0)$ decreases with respect to $4qI_S$ in accordance with the
suppression factor $\varepsilon_F/k_BT$ related to Fermi correlations at
the reservoirs.\cite{buttiker92,kuhn92a} Here, as known, only carriers
around the Fermi level contribute to the noise. As shown by the figure,
the agreement between the results of the Monte Carlo simulation and the
analytical expectations in the nondegenerate  and degenerate limits is
excellent, thus indicating that the carrier injecting statistics achieved
with the proposed model is valid in both regimes.
\par
Figure \ref{cideg} reports the current autocorrelation function $C_I(t)$
in the same structure of Fig.~1 when $\varepsilon_F/k_BT=100$ for several
applied voltages as a function of time normalized to the transit time at
the Fermi level $\tau_T=L(m/2\varepsilon_F)^{1/2}$. The corresponding
$I-U$ curve is shown in the upper inset of Fig.~\ref{cideg}. Due to the
unbalance in the number of carriers reaching the opposite contact, the
current increases linearly with the applied voltage until
$qU=\varepsilon_F$, when the current reaches the saturation value
$I_S=2q\varepsilon_F/h$. For higher $U$ all the carriers injected at the
left lead reach the opposite contact, while no electron injected at the
anode reaches the cathode, and therefore the current saturates. The
conductance in the linear region corresponds to the value of the
fundamental unit $2q^2/h$.\cite{landauer57} The current autocorrelation
function $C_I(t)$ exhibits the following features. At equilibrium it shows
the typical triangular shape, vanishing at $t=\tau_T$ since the
contributions of carriers moving in both directions are symmetrical. This
shape parallels that of a vacuum tube with a constant velocity emitter.
Here, the same shape comes from Pauli principle which allows only carriers
in a small range around the Fermi energy, and therefore moving with
practically the same velocity, to contribute to the noise. When a voltage
below $\varepsilon_F/q$ is applied to the structure, $C_I(t)$ exhibits a
two slope behavior because now the transit times of carriers moving in
opposite directions are different. At voltages higher than
$\varepsilon_F/q$ a negative part appears in $C_I(t)$ since the carriers
injected against the electric field no longer reach the cathode and return
to the anode. At further increasing voltages the negative part appears
sooner due to the shorter time it takes to the carriers to return back to
the right contact.\cite{kuhn92a}
%\par
The low-frequency spectral density $S_I(0)$ as a function of $U$ is shown
in the lower inset of the same figure. At equilibrium the value obtained
corresponds to $S_I(0)=8q^2k_BT/h$, which, when compared with the Nyquist
formula $S_I(0)=4k_BTG$, provides again for the static conductance
$G=2q^2/h$. The fact that in our model the conductance obtained  from the
$I-U$ curve reproduces the fundamental unit value is a valid check of the
correct use of the one-dimensional density of states at the contacts.
Furthermore, the same value $2q^2/h$ is also obtained from noise results
at equilibrium, which proves that the injecting statistics model under
degenerate conditions here proposed is also correct. The voltage
dependence of $S_I(0)$ exhibits a step-like behavior, taking the
equilibrium value up to $U=\varepsilon_F/q$ and half this value for higher
$U$. This behavior is understood as follows. For $U<\varepsilon_F/q$,
carriers around the Fermi level which are injected at both contacts reach
the opposite side and therefore both contribute to the low-frequency
noise. On the contrary, for $U>\varepsilon_F/q$ only carriers injected at
the cathode reach the anode and thus the value of $S_I(0)$ is halved. We
remark that all the results shown in Fig.~\ref{cideg} are in excellent
agreement with previous analytical results in degenerate
systems.\cite{kuhn92a,kuhn92b,reggiani95}
\par
One of the advantages of using particle simulations for the noise analysis
is the possibility to interpret the time and frequency behavior of
fluctuations in terms of different contributions. Thus, in
Fig.~\ref{cidesc}, in the case of degenerate conditions and
$qU/\varepsilon_F=1.01$, $C_I(t)$ is decomposed into velocity $C_V(t)$,
number $C_N(t)$, and velocity-number $C_{VN}(t)$
contributions.\cite{shot1} Here it can be observed that the origin of the
negative part in $C_I(t)$ comes from the velocity-number correlation,
which at zero time exactly compensates the velocity contribution.
Furthermore, from the time dependence of fluctuations three different
characteristic times can be identified. The shortest one corresponds to
the transit of carriers injected from the left contact, as better
evidenced in $C_N$ and $C_{VN}$. It is close to the transit time at
equilibrium $\tau_T$, but slightly shorter due to the acceleration of the
field. A second one is the time taken by the carrier injected at the anode
to reverse its velocity, and is reflected mainly in $C_{VN}$. Finally, the
longest one is the time at which all correlation functions vanish. It
corresponds to the time spent by the electrons injected at the anode to
return back to the same contact.
\par
As final result, in Fig.~\ref{tc} we report the spectrum of the thermal
conductivity $\kappa(f)$ at equilibrium, calculated according to Ref.
\onlinecite{greiner97}. Here, we remark that not only the correlations of
electrical current fluctuations are involved, but also those of the heat
flux and the cross-correlations between both. The oscillatory structure of
Re$[\kappa(f)]$, with geometrical resonances at the inverse of $\tau_T$,
is associated with the fact that all the involved correlation functions
exhibit the triangular shape already found in the case of the current (see
Fig.~\ref{cideg}). A corresponding structure is detectable also in the
imaginary part Im$[\kappa(f)]$. Again, the results of the Monte Carlo
simulation are in excellent agreement with analytical results. In
particular, they reproduce with great accuracy the fundamental unit of
thermal conductance $K=2 \pi^2 k_B^2 T / 3h$ inferred in Ref.
\onlinecite{greiner97}, which again confirms the validity of the proposed
model for the injection statistics at the reservoirs.
\par
In summary, we have presented an injection scheme of electrons at thermal
reservoirs for particle simulations of mesoscopic conductors which takes
into account the binomial distribution of the injected electrons imposed
by Fermi statistics. The model has been validated for the case of
ballistic transport in quasi one-dimensional degenerate conductors. In
particular, at thermal equilibrium we have reproduced the fundamental
units of electrical and thermal conductances, analytically calculated from
the correlation-function formalism. The scheme can be applied continuously
from classical to completely degenerate conditions, and it can be extended
to two and three dimensions, and multi-subband systems. The proposed
scheme is open to applications involving degenerate diffusive conductors.
\par
The authors acknowledge helpful discussions with Prof. A. Reklaitis and
Dr. O. M. Bulashenko. This work has been partially supported by the
Direcci\'on General de Ense\~{n}anza Superior e Investigaci\'on through
the project PB97-1331, and the Physics of Nanostructures project of the
Italian Ministero dell' Universit\'a e della Ricerca Scientifica e
Tecnologica (MURST).
\vspace{-0.5cm}

\begin{figure}
\caption{Low-frequency current spectral density at equilibrium normalized
to $2qI_S$ as a function of the degeneracy factor $\varepsilon_F/k_BT$, at
$T=300 \ K$. Symbols correspond to Monte Carlo calculations and lines to
nondegenerate (solid) and degenerate (dotted) conditions, respectively.
}\label{sieq}
\end{figure}

\begin{figure}
\caption{Current autocorrelation function normalized to the zero-time
value for several applied voltages. The calculations correspond to
degenerate conditions with $\varepsilon_F/k_BT=100$. The upper inset
reports the $I-U$ characteristic of the structure, while the lower inset
shows the $S_I(0)-U$ characteristic. Results are normalized to appropriate
units. Solid lines in the insets correspond to analytical results and full
circles to Monte Carlo calculations.}\label{cideg}
\end{figure}

\begin{figure}
\caption{Decomposition of the current autocorrelation function into
velocity, number and velocity-number contributions for the case of
$qU/\varepsilon_F=1.01$ and $\varepsilon_F/k_BT=100$.}\label{cidesc}
\end{figure}

\begin{figure}
\caption{Spectrum of thermal conductivity at equilibrium normalized to the
theoretically predicted zero-frequency value. Solid and dotted lines
correspond to the real and imaginary parts, respectively. Calculations are
performed for $\varepsilon_F/k_BT=100$.}\label{tc}
\end{figure}


\begin{thebibliography}{99}
%
\bibitem{kogan96}
\vspace{-1.2cm} Sh. Kogan, {\it Electronic Noise and Fluctuations in
Solids} (Cambridge University Press, Cambridge, 1996).

\bibitem{dejong97}
M. J. M. de Jong and C. W. J. Beenakker, in {\it Mesoscopic Electron
Transport}, Vol. 345 of {\it NATO Advanced Study Institute Series E:
Applied Science}, edited by L. P. Kowenhoven, G. Sch\"on, and L. L. Sohn
(Kluwer, Dordrecht, 1997), p. 225.

\bibitem{landauer57}
R. Landauer, IBM J. Res. Dev. {\bf 1}, 223 (1957).

\bibitem{buttiker90}
M. B\"uttiker, Phys. Rev. Lett. {\bf 65}, 2901 (1990).

\bibitem{martin92}
Th. Martin and R. Landauer, Phys. Rev. B {\bf 45}, 1742 (1992).

\bibitem{buttiker92}
M. B\"uttiker, Phys. Rev. B {\bf 46}, 12485 (1992).

\bibitem{kuhn92a}
T. Kuhn, L. Reggiani, and L. Varani, Superlatt. Microstruct. {\bf 11}, 205
(1992).

\bibitem{kuhn92b}
T. Kuhn and L. Reggiani, Nuovo Cimento D {\bf 14}, 509 (1992).

\bibitem{reggiani95}
L. Reggiani, L. Varani, T. Gonz\'alez, D. Pardo, and T. Kuhn, in {\it
Proc. 22nd Int. Conf. on the Physics of Semiconductors}, edited by D. J.
Lockwood (World Scientific, Singapore, 1995), p. 1963.

\bibitem{greiner97}
A. Greiner, L. Reggiani, T. Kuhn, and L. Varani, Phys. Rev. Lett. {\bf
78}, 1114 (1997).

\bibitem{nagaev92}
K. E. Nagaev, Phys. Lett. A {\bf 169}, 103 (1992).

\bibitem{naveh97}
Y. Naveh, D. V. Averin, and K. K. Likharev, Phys. Rev. Lett. {\bf 79},
3482 (1997).

\bibitem{suk98}
E. V. Sukhorukov and D. Loss, Phys. Rev. Lett. {\bf 80}, 4959 (1998).

\bibitem{liu97}
R. C. Liu, P. Eastman, and Y. Yamamoto, Solid State Commun. {\bf 102}, 785
(1997).

\bibitem{shot1}
T. Gonz\'alez, O. M. Bulashenko, J. Mateos, D. Pardo, and L. Reggiani,
Phys. Rev. B {\bf 56}, 6424 (1997).

\bibitem{shot3}
O. M. Bulashenko, J. Mateos, D. Pardo, T. Gonz\'alez, L. Reggiani, and J.
M. Rub\'{\i}, Phys. Rev. B {\bf 57}, 1366 (1998).

\bibitem{shot4}
T. Gonz\'alez, C. Gonz\'alez, J. Mateos, D. Pardo, L. Reggiani, O. M.
Bulashenko, and J. M. Rub\'{\i}, Phys. Rev. Lett. {\bf 80}, 2901 (1998).

\bibitem{shot5}
T. Gonz\'alez, J. Mateos, D. Pardo, O. M. Bulashenko, and L. Reggiani,
Semicond. Sci. Technol. {\bf 13}, 714 (1998).

\bibitem{jacoboni89}
C. Jacoboni and P. Lugli, {\it The Monte Carlo Method for Semiconductor
Device Simulation} (Springer, Berlin, 1989).

\bibitem{varani94}
L. Varani, L. Reggiani, T. Kuhn, T. Gonz\'alez, and D. Pardo, IEEE Trans.
Electron Devices {\bf ED-41}, 1916 (1994).

\bibitem{woolard94}
D. L. Woolard, H. Tian, M. A. Littlejohn, and K. W. Kim, IEEE Trans.
Electron Devices {\bf ED-41}, 601 (1994).

\bibitem{gonzalez96}
T. Gonz\'alez and D. Pardo, Solid-State Electron. {\bf 39}, 555 (1996).

\bibitem{levitov93}
L. S. Levitov and G. B. Lesovik, Pis'ma Zh. Eksp. Teor. Fiz. {\bf 58}, 225
(1993) [JETP Lett. {\bf 58}, 230 (1993)]; L. S. Levitov, H. Lee, and G. B.
Lesovik, J. Math. Phys. {\bf 37}, 4845 (1996).

\bibitem{remark1}
In this sense the injection is uniform in time, since electrons are
injected equally spaced in time, which can also be interpreted as an
injection {\it periodic} in time, with period $1 / \zeta_{k_i,\Delta k}$.

\end{thebibliography}
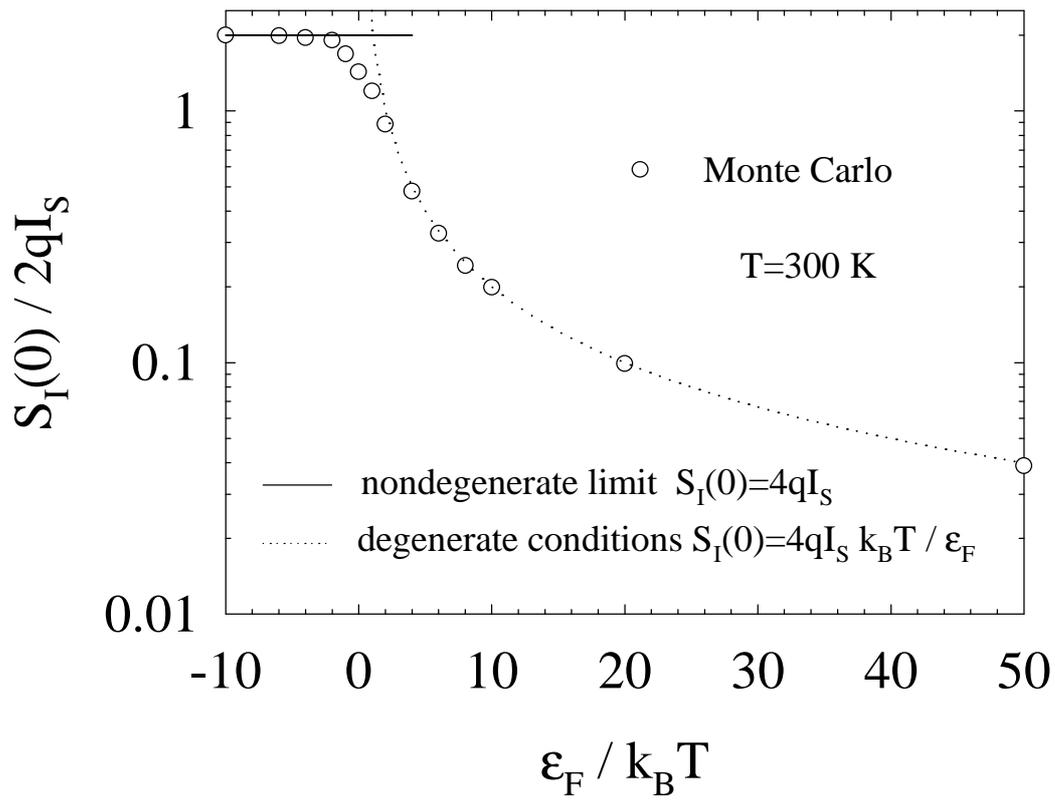
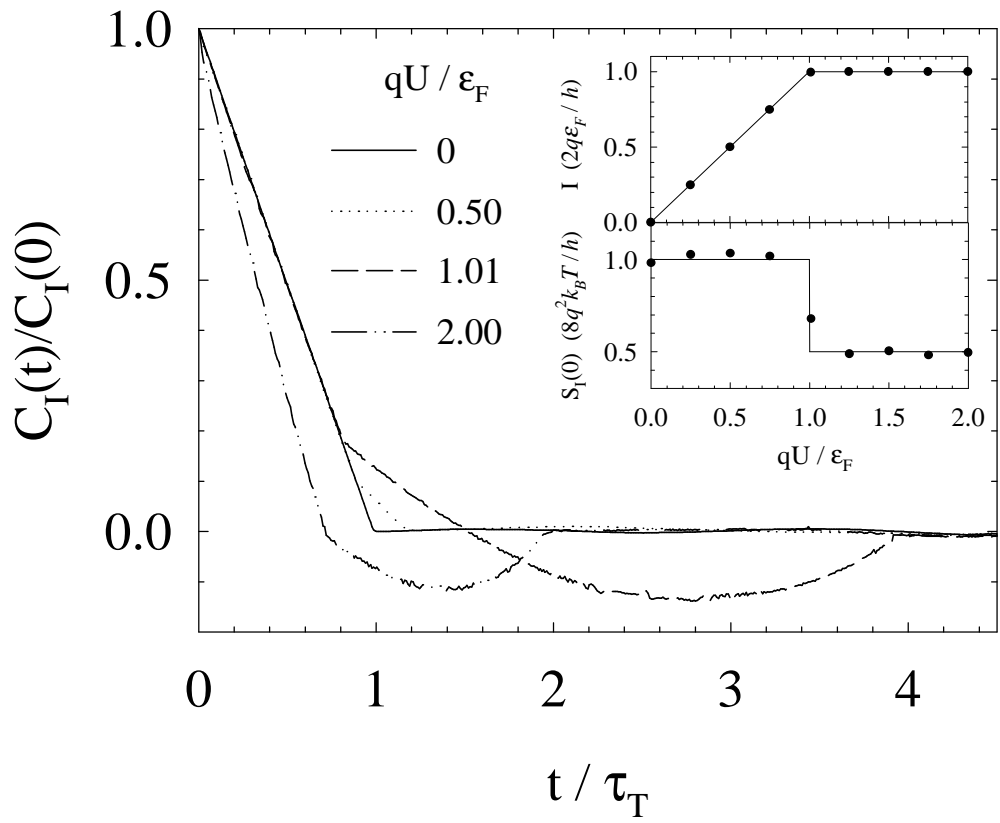
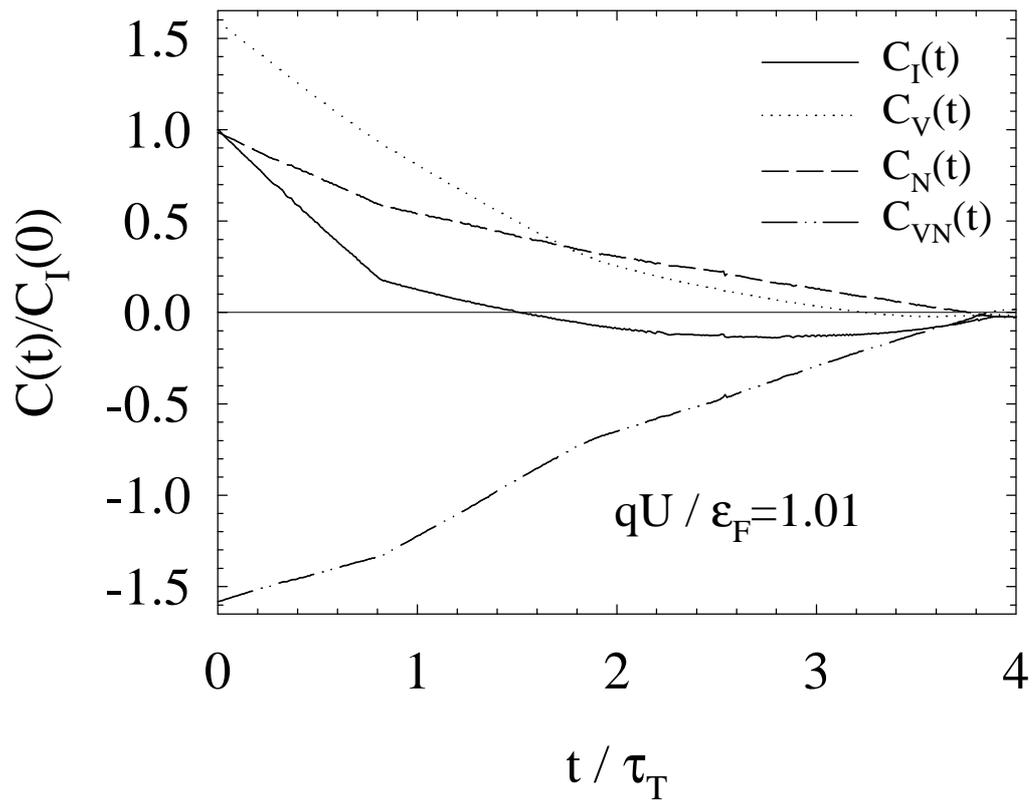
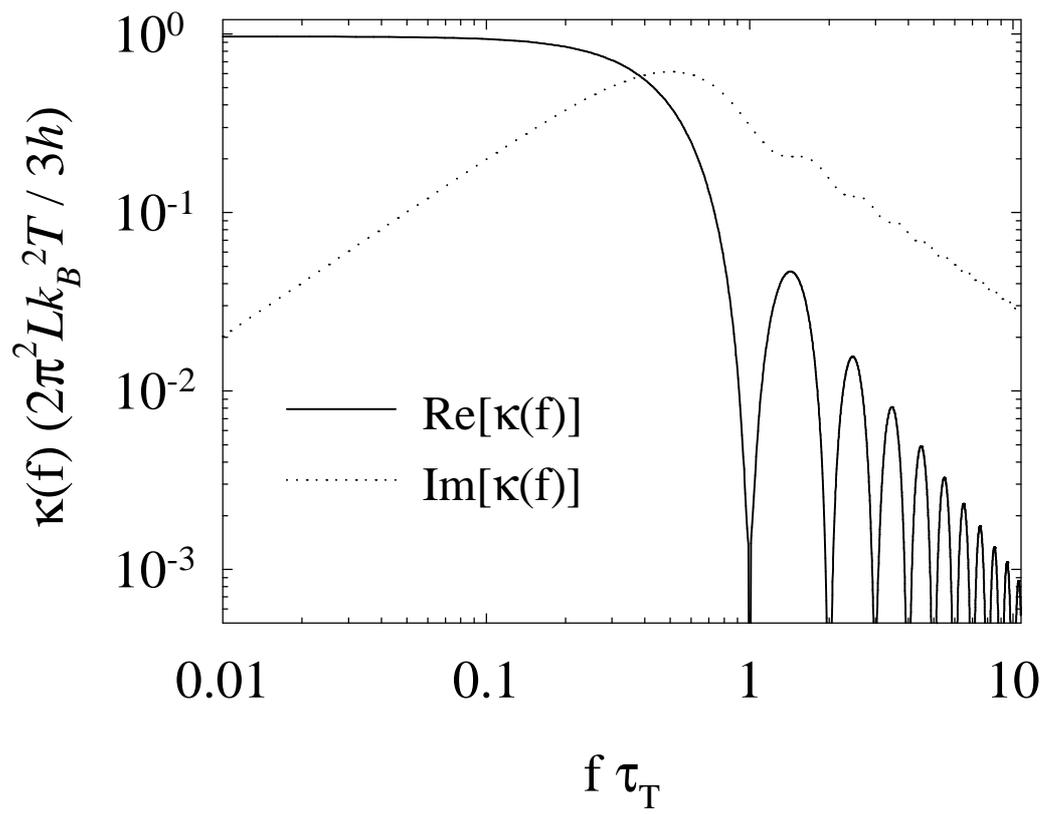
\end{document}